\documentclass{pnastwo}

\usepackage{graphicx}
\usepackage{amssymb,amsfonts,amsmath}
\usepackage{amsmath}

\def\fs{\footnotesize}

\url{www.pnas.org/cgi/doi/10.1073/pnas.0709640104}
\copyrightyear{2008}
\issuedate{Issue Date}
\volume{Volume}
\issuenumber{Issue Number}

\begin{document}

\title{Extraterrestrial sink dynamics in granular matter}

\author{E. Altshuler\affil{1}{``Henri Poincar\'{e}" Group of Complex Systems, Physics Faculty, University of Havana, 10400 Havana, Cuba},
H. Torres \affil{1}{``Henri Poincar\'{e}" Group of Complex Systems and Superconductivity Laboratory, Physics Faculty-IMRE, University of Havana, 10400 Havana, Cuba},
A. Gonz{\'a}lez-Pita\affil{1}{``Henri Poincar\'{e}" Group of Complex Systems and Superconductivity Laboratory, Physics Faculty-IMRE, University of Havana, 10400 Havana, Cuba},
G. S{\'a}nchez-Colina \affil{1}{``Henri Poincar\'{e}" Group of Complex Systems and Superconductivity Laboratory, Physics Faculty-IMRE, University of Havana, 10400 Havana, Cuba},
C. P\'{e}rez-Penichet\affil{1}{``Henri Poincar\'{e}" Group of Complex Systems and Superconductivity Laboratory, Physics Faculty-IMRE, University of Havana, 10400Havana, Cuba},
S. Waitukaitis \affil{2}{``James Franck" Institute, University of Chicago, U.S.A.},
R. C. Hidalgo \affil{3}{Department of Physics and Applied Mathematics, University of Navarra, Pamplona, Spain}}

\contributor{Submitted to Proceedings of the National Academy of Sciences
of the United States of America}

\maketitle
\begin{article}
\begin{abstract}A loosely packed bed of sand sits precariously on the fence between mechanically stable and flowing states.  This has especially strong implications for animals or vehicles needing to navigate sandy environments, which can sink and become stuck in a ``dry quicksand'' if their weight exceeds the yield stress of this fragile matter.  While it is known that the contact stresses in these systems are loaded by gravity, very little is known about the sinking dynamics of objects into loose granular systems under gravitational accelerations different from the Earth's ($g=9.8$ m/s$^2$).  A fundamental understanding of how objects sink in different gravitational environments is not only necessary for successful planetary navigation and engineering, but it can also improve our understanding of celestial impact dynamics and crater geomorphology.  Here we perform and explain the first systematic experiments of the sink dynamics of objects into granular media in different gravitational accelerations $g_{eff}$.  By using an accelerating experimental apparatus, we explore conditions ranging from $0.4g$ to $1.2g$.  With the aid of discrete element modeling simulations, we reproduce these results and extend this range to include objects as small as asteroids and as large as Jupiter.  Surprisingly, we find that the final sink depth is \emph{independent} of $g_{eff}$, an observation with immediate relevance to the design of future extraterrestrial structures land-roving spacecraft.  Using a phenomenological equation of motion that includes a gravity-loaded frictional term, we are able to quantitatively explain the experimental and simulation results.
\end{abstract}


\dropcap{V}irtually all exploration and development of extraterrestrial settings involves navigation in and on loose
granular media.  This is due in large part to the fact that in the
the geomorphology of these non-Earth environments is dominated by
wind or gravity driven granular flows, which create sandy
environments ranging from the pebbles and sand on the surface of
Mars \cite{Shinbrot2003, Almeida2008} to the loosely packed dust on
asteroids \cite{Thomas2005, Miyamoto2007}.  While interest in the
exploration and utilization of these environments has never been
higher, little is known about how objects behave in and on granular
media with gravitational accelerations other than $g$.  Even
seemingly simple and common phenomena, such as the sinking of an
object set at rest on the free granular surface, are not understood.
Nonetheless, a fundamental understanding of how objects penetrate
and sink in different gravitational environments is fundamental to
the success of extraterrestrial engineering and navigation.  As a
case in point, such an understanding may have helped prevent the
difficulties encountered by the Mars rover, Spirit, as it sank into
and tried to escape from a sand dune in 2009 \cite{NASA2009}.  More
complicated future space endeavors, such as asteroid or lunar mining
\cite{Elvis2012}, will certainly involve both navigation and
construction on granular surfaces, and knowing how objects settle in
these environments is a critical first step.

For Earth-like conditions, the complexity of the motion on and sinking into granular surfaces is well-studied \cite{Uehara2003,Walsh2003,Boudet2006,Vet2007,Katsuragi2007,Pacheco2011,Katsuragi2012,Kondic2012, Ruiz-Suarez2013}.  A handful of attempts have tried to resemble, albeit in indirect ways, low gravity conditions \cite{Goldman2008,Brzinski2010,Brzinski2013,Constantino2011}, but most studies have instead focused on the role of packing fraction, grain density, or impactor velocity.  For locomotion, the importance of these parameters is particularly well-illustrated by the work of Li {\it et al.} \cite{Li2009}, who demonstrated the extreme sensitivity of the motion of a legged robot to the packing fraction of the granular soil and the strong dependence of its step size on the depth of penetration of the legs in the sand.

Here we systematically study the sink dynamics of a sphere into loose granular material at different gravitational accelerations, both above and below that of Earth.  By conducting experiments in an accelerating lab frame, we subject a sinking sphere to gravitational accelerations, $g_{eff}$, similar to those those at the surfaces of Mars, Venus, Earth, Neptune and Uranus.  While we confirm the previously reported (but hereto unexplained) result that the total sinking time scales as the ${g_{eff}}^{-1/2}$, we also find, surprisingly, that the final sink depth is independent of $g_{eff}$.  We confirm and extend these experimental results to gravitational accelerations encountered in asteroids and heavier planets with the aid of 3D discrete element modeling (DEM) simulations.  We interpret and explain the observations quantitatively with a phenomenological equation of motion that explicitly relates the observed force to grain-grain contact loading via gravity.

\begin{figure}[h]
\includegraphics[scale=0.9]{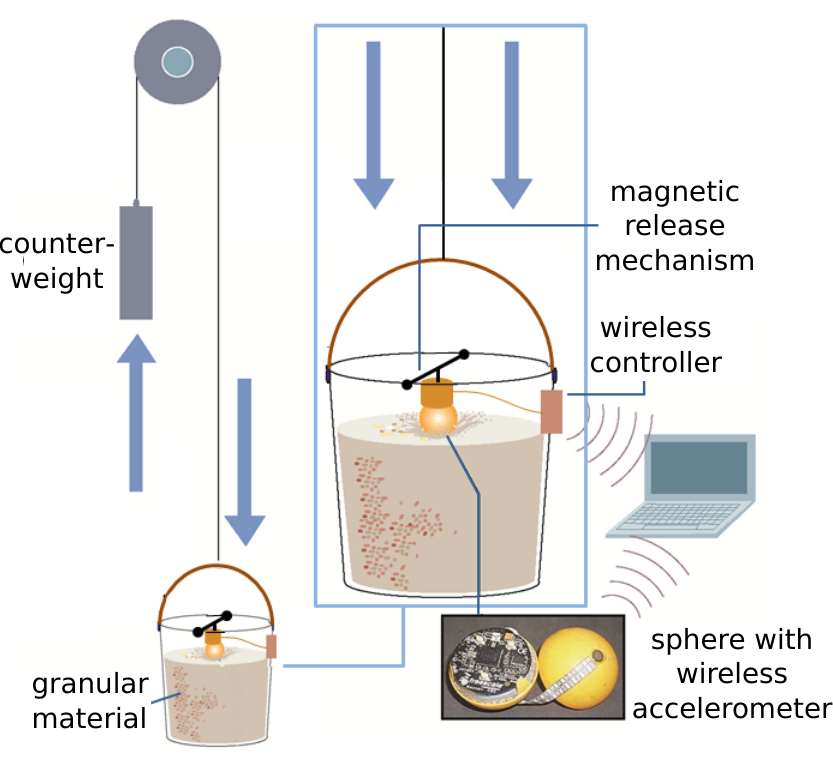}
\caption{Experimental setup. A $15$~m tall Atwood machine controls
the downward/upward acceleration of a 30 cm diameter laboratory in a
bucket filled with the granular medium (expanded polystyrene beads),
shown in detail.  As the bucket falls/rises, a sphere is released
from rest and allowed to sink while ``feeling'' the effective
gravity $g_{eff}$ of the accelerating frame. The magnetic release
mechanism of the sphere is controlled remotely and initiated once an
accelerometer attached to the bucket indicates stable acceleration.
We use a second accelerometer embedded into the sphere for real time
measurement of the post-release acceleration (and, after
integration, velocity and position).} \label{fig:fig1}
\end{figure}

\section{Results and discussion}
\subsection{Experimental results}

We vary $g_{eff}$ using a 15 m tall Atwood machine in which one of the counterweights is a wireless granular laboratory (Fig.~1).  The laboratory consists of a cylindrical bucket ($30$ cm diameter by $26$ cm depth) filled with expanded polystyrene beads (average diameter $\sim$5 mm).  When the bucket is allowed to rise or fall, the granular media and the equipment inside it ``feel", for a few seconds, a gravitational acceleration $g_{eff}$ different than $g$ (larger if the bucket is rising and smaller if it is falling).  As the bucket rises/falls, we release a sphere held at rest just above the free surface and let it sink into the granular medium.  The sphere houses a three-axis wireless accelerometer in its interior, which allows us to record its instantaneous acceleration in real time.  The total mass of the impactor/accelerometer is $m=23$~g, light enough to prevent the ``infinite penetration'' encountered by Pacheco \emph{et al.} \cite{Pacheco2011}.

In Fig.~2({\it A}) we plot the time-dependence of the sphere's
acceleration $a$ vs.~time $t$ relative to the bucket (normalized by
Earth's gravity $g=$~9.8~m/s$^2$) for six representative values of
$g_{eff}$ (note we define positive acceleration as pointing
downward).  Each curve has three well defined sections: (i) an
initial region of positive slope corresponding to the release of the
sphere (this is caused by magnetic forces of the release mechanism
and occurs in the first $\sim$50 ms), (ii) a second region of
negative slope, where most of the penetration process takes place,
and (iii) a third region of positive slope which includes the sudden
stopping of the sphere (note this sudden arrest was also seen in
previous experiments performed at $g_{eff}=g$).  In contrast with
most studies in the literature, our ultralight granular medium
permits us to observe the initial positive segment during which the
frictional forces on the granular medium are small.  We also note
the presence of a brief, damped oscillation that occurs near the
instant the sphere comes to rest, which suggests a ``shock'' against
a jammed granular ``wall" (as shown later, this feature is
reproduced in our DEM simulations).  The
oscillations in the $a$ vs. $t$ curves are not symptomatic of the
resolution of our accelerometer but instead are real fluctuations
from the granular medium (we confirm this in the simulation results,
which show similar oscillations).

\begin{figure}
\centering
\includegraphics[scale=0.85]{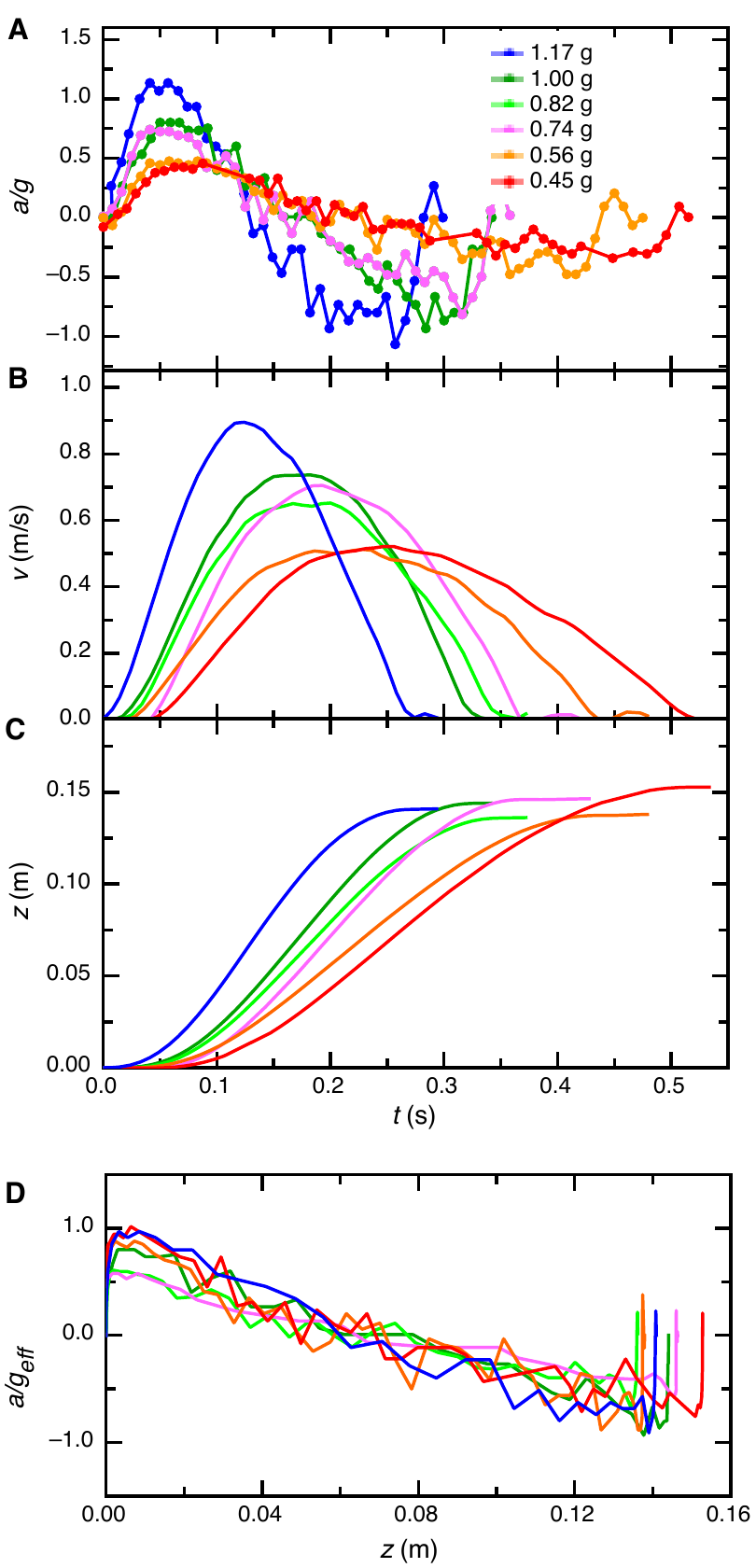}
\caption{Granular sink dynamics in experiment.  (A)  Sphere acceleration \textit{a/g} vs. time \textit{t} for six representative values of $g_{eff}$.
Values for $g_{eff}$ are indicated in the legend in the figure and also correspond to following panels.
(B)  Time dependence of the sphere's velocity $v$ via numerical integration of (A).
(C)  Time dependence of sphere's penetration below free granular surface $z$, calculated via integration of (B).
(D) Normalized sphere acceleration $a/g_{eff}$ vs. penetration distance $z$.}
\label{fig:fig2}
\end{figure}

Comparison of the different curves in Fig.~2({\it A}) shows that as $g_{eff}$ decreases from the blue curve (near Uranus' gravity) to the red curve (near Mars' gravity), the peak acceleration increases while the depth of the minimum decreases.  What's more, the duration of the process as a whole increases.  This point is made particularly clear in Fig.~2({\it B}), where we integrate $a$ vs. $t$ and plot the velocities $v$ vs. $t$, which also shows that the maximum speed of the sphere increases with higher $g_{eff}$.  In Fig.~2({\it C}), we integrate once more to plot the distance travelled below the surface $z$ vs. $t$, which reveals the key observation that the final penetration depth $z_{sink}$ is essentially the same for all $g_{eff}$ (average value $z_{sink}=14.0\pm0.6$ cm).  If instead of plotting the acceleration vs. $t$ we plot it against $z$ (normalized by $g_{eff}$, as in Fig.~2({\it D})), we collapse the data to a line with slight upward curvature (apart from the brief initial and final moments, corresponding to sections (i) and (iii), respectively).

\subsection{Simulation results}

Figure~3({\it A-D}) shows the corresponding simulation results for
similar gravitational accelerations to those used in the
experiments.  With the exception of the brief time period in which
the magnet turns off (which we did not simulate), the simulation
results are strikingly similar to the experiments.  Several
experimental features are reproduced quantitatively with no free
parameters, {\it e.g.} the duration of the penetration process, the size
of the acceleration peaks, the maximum velocities and the final
penetration depth.  Even the sudden drop to zero acceleration and
the acceleration fluctuations (which could have been interpreted as
noise) are present.  Closer inspection here also reveals that the
fluctuations in the acceleration become stronger as the sphere motion comes closer to stopping.  This behavior may be associated with the building
up and breaking down of force chains during the penetration process,
suggesting the stopping and eventual static support of the intruder
is associated with the medium transitioning from a fluidized to
jammed state \cite{Kondic2012, Bi2011}.

\begin{figure}
\centering
\includegraphics[scale=0.85]{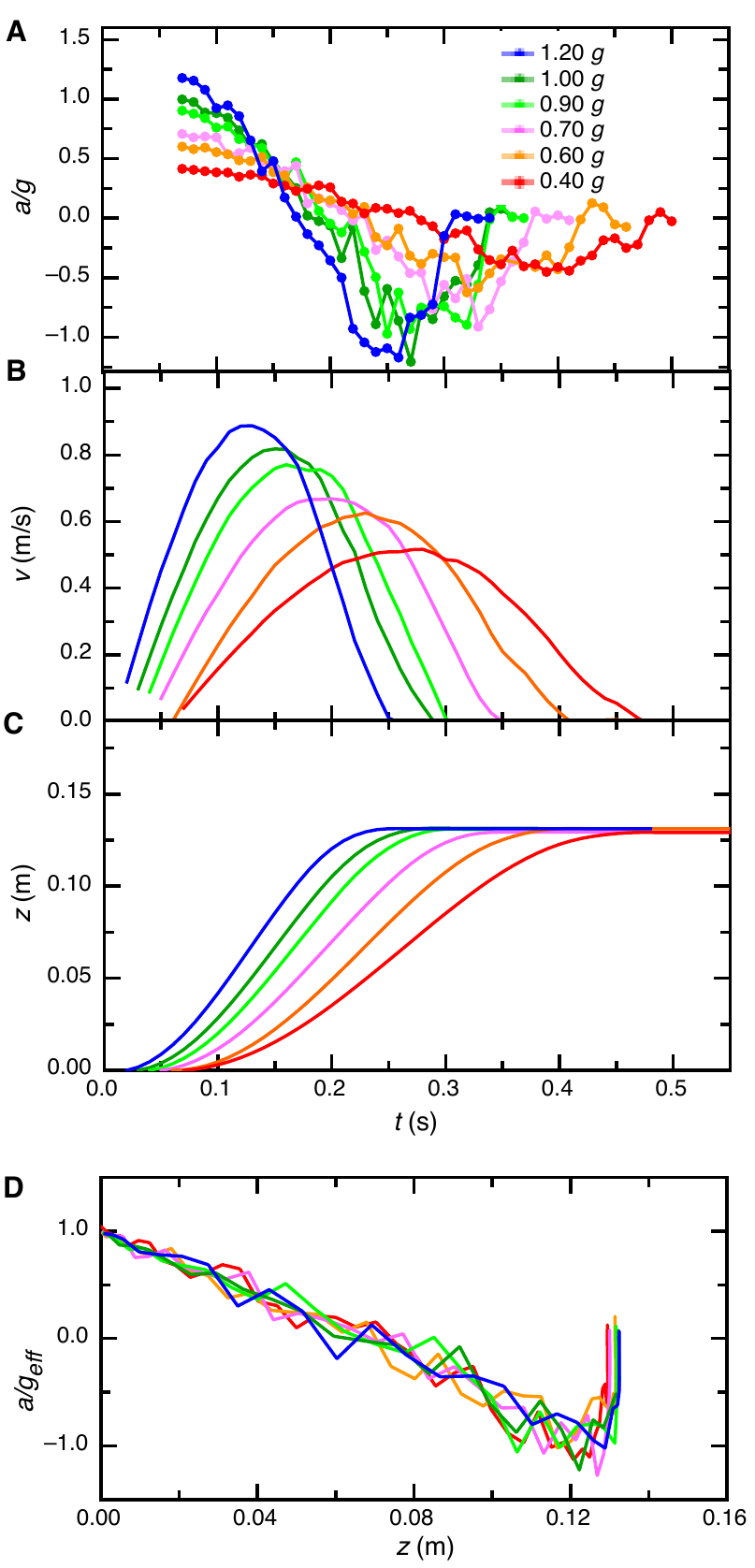}
\caption{Granular sink dynamics in 3D DEM simulations.  (A)  Sphere acceleration \textit{a/g} vs. time \textit{t} for six representative values of $g_{eff}$.  Values for $g_{eff}$ are indicated in the legend in the figure and also correspond to following panels.  (B)  Time dependence of the sphere's velocity $v$ via numerical integration of (A).  (C)  Time dependence of sphere's penetration below free granular surface $z$, calculated via integration of (B).  (D) Normalized sphere acceleration $a/g_{eff}$ vs. penetration distance $z$.}
\label{fig:fig3}
\end{figure}

\subsection{Equation of motion}

In Fig.~4, we present the combined experimental and simulation results for the final sink depth $z_{sink}$ (Fig.~4(A)) and sink time $t_{sink}$ (Fig.~4(B)), which shows that $z_{sink}$ is essentially independent of $g_{eff}$ while $t_{sink}$ scales like $g^{-1/2}_{eff}$.  The fact that the penetration depth of the sinking sphere is independent of $g_{eff}$ is both surprising and puzzling.  To begin to explain it, we look to the work of Pacheco-V{\'a}zquez {\it et al.} \cite{Pacheco2011}, who proposed a simple equation of motion for an object impacting into a granular medium,
\begin{equation}\label{eq:Pacheco}
 ma=mg-\eta v^2-\kappa \lambda (1-e^{-(z/\lambda)}),
\end{equation}
where $m$ is the impactor mass, $\eta$ characterizes any intertial
drag, $\kappa$ is a friction-like coefficient related to the pressure in the granular medium (units [N/m]), and
$\lambda$ is a characteristic length of the order of the width of
the container holding the medium.  (The exponential term arises from
the well-known Janssen effect in which the pressure in a granular
system saturates at a finite depth owing to redistribution of weight
to the container walls \cite{Janssen1895}.)  In recent experiments
of a sphere falling through a tall silo of polystyrene beads at
Earth gravity, it was found that the combination of the inertial
drag term ($\propto v^2$) and the saturating, depth dependent
pressure term ({\it i.e.} the exponential) can lead to ``infinite
penetration'' of the projectile at a constant, terminal velocity if
its mass is sufficiently large.  Here, however, because the sphere
starts at rest and is relatively light, we can ignore the drag term
(the maximum velocity we see is $\sim1$ m/s and the small values of
$\eta$ involved \cite{Pacheco2011,Torres2012} put the term $\eta
v^2$ approximately one order of magnitude smaller than the other terms
during the whole penetration process). Thus, the equation of motion
can be approximated as
\begin{equation}\label{eq:PachecoSmallV}
 ma=mg_{eff}-\kappa \lambda (1-e^{-(z/\lambda)}).
\end{equation}
This proposed form quickly explains the shape of the $a/g_{eff}$ vs.
$z$ curves shown in Fig.~2({\it D}) (and in particular reproduces
the slight upwards curvature).  By fitting each of the $a/g_{eff}$
vs. $z$ curves with Eq.~\ref{eq:PachecoSmallV} while using the bucket
radius for the parameter $\lambda$ and leaving $\kappa$ as a free
parameter, we find the linear relationship $\kappa = \alpha
g_{eff}$, where $\alpha=0.415 \pm 0.004$~Ns$^2/$m$^2$, as shown in
Fig.4({\it C}).  This linear relationship has been proposed before
(see, for example, \cite{Katsuragi2007}), but here we demonstrate
its validity at different gravities for the first time. Other
experiments that have imitated effective gravities (by flowing air
upwards through the grains \cite{Brzinski2010,Brzinski2013} or with
granular-liquid mixtures \cite{Constantino2011}) have interpreted
this relationship as arising from the loading of frictional forces in
grain-grain contacts inside the granular media, and our findings explicitly
confirm this is the case.

Beyond being consistent with the $a$ vs.~$z$ curves from the experiments and simulations, this model also predicts the observation that the total sink time scales like $g_{eff}^{-1/2}$.  To show this, we begin by rewriting Eq.~\ref{eq:PachecoSmallV} as
\begin{equation}\label{eq:Scott}
v\frac{dv}{dz}=g_{eff}\bigg{[}1-\frac{\alpha \lambda}{m}\bigg{(}1-e^{-(z/\lambda)}\bigg{)}\bigg{]}.
\end{equation}
We integrate this equation with respect to $z$ (with the initial conditions $z_0=0$ and $v_0=0$) to find
\begin{equation}
 \label{eq:Scott2}
 \frac{1}{2}v^2=g_{eff}\bigg{[}  z\bigg{(}1-\frac{\alpha \lambda}{m}\bigg{)} -\frac{\alpha \lambda^2}{m}\bigg{(} e^{-z/\lambda}-1\bigg{)}   \bigg{]}.
\end{equation}
Next, we isolate the velocity term, take the square root of both sides (note we are interested in $t_{sink}>0$ and thus use the positive root), and integrate once more, which gives
\begin{equation}\label{eq:t_sink}
    t_{sink}=(2g_{eff})^{-1/2} \int_0^{z_{sink}}\bigg{[}1-\frac{\alpha \lambda}{m}\bigg{(}1-e^{(-z/\lambda}\bigg{)}\bigg{]}^{-1/2}\,dz.
\end{equation}
The term in the integral is independent of $g_{eff}$ (as $\lambda$, $\alpha$, and $m$ strictly independent of $g_{eff}$ and $z_{sink}$ is empirically so).  Consequently, we conclude that $t_{sink}\propto g_{eff}^{-1/2}$.

Finally, we can also show that the final sink depth of the sphere, $z_{sink}$, is independent of $g_{eff}$.  When the sphere reaches its resting spot, the velocity vanishes and thus we can set the lefthand side of Eq.~\ref{eq:Scott2} to zero, \textit{i.e.}
\begin{equation} 
  \label{eq:z_sink}
  \bigg{(} 1-\frac{\alpha \lambda}{m} \bigg{)}z_{sink} = \frac{\alpha \lambda^2}{m} \bigg{(} e^{-z_{sink}/\lambda} -1  \bigg{)}.
\end{equation}
This is a well-known transcendental equation that cannot be solved analytically.  However, quick inspection of it reveals immediately that $z_{sink}$ is \textit{independent} of the gravitational acceleration as $g_{eff}$ does not appear anywhere in the equation.  Though we can't get an analytic solution for $z_{sink}$, we can use the experimental parameters $\lambda=0.15$ m and $\alpha=0.415$ Ns$^2$/m$^2$ to solve Eq.~\ref{eq:z_sink}  numerically for our system, which gives $z_{sink} \approx 0.15$ m,  close to what we actually measure.  The fact that it is somewhat larger may result from ignoring the velocity term in Eq.~\ref{eq:Pacheco}, which would tend to make the sphere stop a little earlier.  (Indeed, numerically solving the the differential equation [Eq.~\ref{eq:Pacheco}] directly with the value for $\eta$ from Pacheco {\it et al.} \cite{Pacheco2011} gives the value $z_{sink}=0.14$ m, in better agreement still with the data from the experiments and simulations.)

\section{Conclusion}

Our work here is the first report on the full penetration dynamics of an object sinking into granular media at different gravitational accelerations.  By using a freely-falling experimental laboratory, we are able to investigate gravitational environments both larger and smaller than that of Earth, ranging roughly from the conditions of Mars to Uranus.  We reproduce and extend the range of these results with the aid of DEM simulations, which highlight the importance of transient force fluctuations in the penetration process that may be related to the continual build up and break down of granular force chains.  In both the experiments and simulations, we make a counter-intuitive observation in the sinking process that has important implications for extraterrestrial navigation and engineering, namely that the final sink depth of an object set at rest on granular media is \emph{independent} of the ambient gravitational acceleration.  We are able to explain this peculiar observation with a force law which includes a depth dependent frictional term that is proportional to $g_{eff}$, which effectively removes any gravitational term from the equation of motion at the point of static equilibrium.  This finding in particular suggests that Earth-based experiments aimed at reproducing the conditions of a robot navigation on or a structure being built on another planet or asteroid should be performed without ``adjusting" the mass of the device for the new gravity conditions.
\begin{figure}
\centering
\includegraphics[scale=0.85]{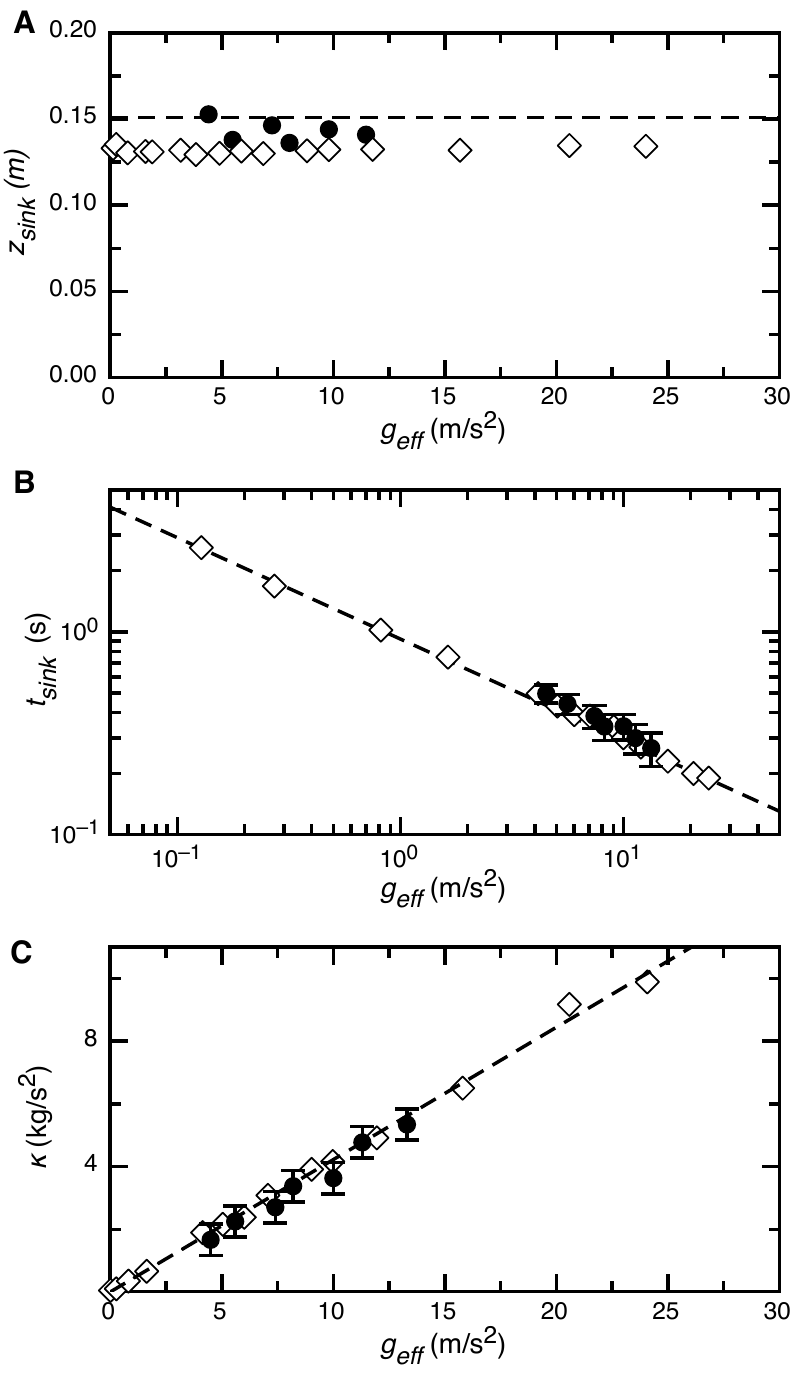}
\caption{ (A) Final penetration depth $z_{sink}$ vs $g_{eff}$ for experiment (black circles) and simulations (open diamonds).  Dashed line is predicted final penetration depth based on Eq.~\ref{eq:z_sink}.  (B)  Dependence of $t_{sink}$ vs. $g_{eff}$.  Fit to simulation data is power law with exponent $-1/2$, as predicted in Eq.~\ref{eq:t_sink}.  (C) Frictional sink parameter $\kappa$ vs. $g_{eff}$.  Values for $\kappa$ are calculated from individually fitting $a$ vs. $z$ curves to Eq.~\ref{eq:Scott} with $\lambda=15$ cm (\textit{i.e.} the radius of the container holding the granular material).  Symbols in (B), (C) are the same as in (A).}
\label{fig:fig4}
\end{figure}

\begin{materials}
\section{Experimental details}

The granular media consists of expanded polystyrene particles with a density of {\fs $0.014 \pm 0.002$} ~g/cc and a diameter distribution ranging from approximately {\fs$2.0$} to {\fs$6.5$} mm (with a peak at {\fs$5.8$}~mm).  To ensure that the system has a similar initial configuration each experiment, we use the following procedure adapted from Torres {\emph {\fs et al.}}~\cite{Torres2012}. First, we inject air from below through a wire mesh with a pressure ramp just until the top of the bed just becomes fluidized.  Then we slowly lower the pressure until there is zero flow.  Next, we shake the container horizontally for {\fs$5$} seconds (the oscillations are approximately sinusoidal, with a period of {\fs$0.225 \pm 0.004$}~s and an acceleration amplitude of {\fs$1.9 \pm 0.3$}~m/s{\fs$^2$}).  This process repeatably produces a volume fraction of {\fs$0.68 \pm 0.01$} and maximum angle of stability of {\fs$30.29^o \pm 0.50^o$}.

The sphere is quickly released into free-fall with the aide of a magnetic latch.  The exact moment of release is determined by remotely observing the bucket acceleration with a computer and, once it is confirmed that the bucket moves with constant acceleration {\fs$g_{eff}$}, deactivating the latch.  Care is taken to ensure that little lateral motion occurs and that at release the bottom of the sphere is just gently touching the free granular surface.  The 3-axis accelerometer inside the sphere has a resolution of {\fs$10^{-4}g$} and is able to transmit data in real time at {\fs $2.4$}~GHz to a USB node on an external PC at a data point rate of {\fs$120$}~Hz.  The device had a saturation acceleration of {\fs$\sim8g$} \cite{Zigbee}.

\section{Simulation Details}

We use discrete element modeling ({\fs$DEM$}) to simulate a large sphere sinking into a granular bed composed of smaller spheres \cite{Poschel2005}.  The implementation is a hybrid {\fs$CPU/GPU$} algorithm that allows us to efficiently evaluate the dynamics of several hundred of thousands of particles \cite{raul2013,sudafrica2013,Owens2008}.  We initiate each simulation by generating a random granular packing of monodisperse spheres (radius {\fs$r$} and density  {\fs$\rho$}) at packing fraction {\fs$\phi=0.62\pm0.02$}.  The spherical intruder ({\fs$R=8r$} and density {\fs$\rho_{int}=50 \rho$}) is released from the free granular surface with zero initial velocity.

For each particle {\fs$i=1...N$}, the {\fs $DEM$} simulation includes three translational degrees of freedom and the rotational movement is described by a quaternion formalism.
In our approach, the normal interaction force between the particles  {\fs ${\vec F}_{ij}^{n}$} depends non-linearly on the particles overlap distance {\fs $\delta$}.  Moreover, the local dissipation is introduced by a non-linear viscous damping term, which depends on the normal relative velocity {\fs${\vec v}_{rel}^{n}$}.  Hence, the total normal force reads as {\fs${\vec F}_{ij}^{n} =- k_n \delta^{3/2} {\hat n} - \gamma_n {\vec v}_{rel}^{n}  \delta^{1/4} $}, where {\fs$k_n$} and {\fs$\gamma_n$} represent elastic and damping coefficients, respectively.  This formulation corresponds to a  non-linear Herzt's contact with
constant restitution coefficient \cite{Poschel2005}. The tangential component {\fs $F_{ij}^{t}$} also includes an elastic term and a viscous term, {\fs ${\vec F}_{ij}^{t} =-k_t {\vec \xi}- \gamma_t  {\vec v}_{rel}^{t}$ },
where {\fs$\gamma_t$} is a damping coefficient and {\fs$ {\vec v}_{rel}^{t}$}  is the tangential relative velocity of the overlapping pair. The variable {\fs$|{\vec \xi}|$} represents the elongation of an imaginary spring with elastic constant {\fs$k_t$}. As long as there is an overlap between the interacting particles, {\fs${\vec \xi}$} increases as {\fs$d{\vec \xi}/dt = {\vec v}_{rel}^t$}  \cite{Poschel2005}.  The elastic tangential elongation {\fs${\vec \xi}$}  is truncated as necessary to satisfy the Coulomb constraint {\fs$|{\vec F}_{ij}^{t}|<\mu |{\vec F}_{ij}^{n}|$}, where {\fs$\mu$} is the friction coefficient.

In all the simulations reported here, the values of the normal elastic and damping coefficients correspond to particles with a Young's modulus {\fs$Y=10^7 Pa$}, normal restitution coefficient {\fs$e_n=0.2$}, friction coefficient {\fs$\mu=0.5$} and density {\fs$\rho=14.0$} kg/m{\fs$^3$} .  We keep {\fs$\frac{k_t}{k_n} = \frac{2}{7}$}, {\fs$\frac{\gamma_t}{\gamma_n}=0.1$} and only modify the gravitational acceleration $g_{eff}$ from one simulation to the next.  For these parameters, the time step was set in {\fs$\Delta t = 10^{-6}s$}. The equations of motion are integrated using a Fincham's leap-frog algorithm (rotational) \cite{Fincham1992} and a Verlet Velocity
algorithm (translational) \cite{Verlet1968}.

\end{materials}

\begin{acknowledgments}
We thank O. Ramos, J. Wu, C. Ruiz-Su{\'a}rez and A. J. Batista-Leyva for material support and useful discussions, and the Zeolites Group (IMRE) for collaboration during the experiments.  The Spanish MINECO Projects FIS2011-26675 and the University of Navarra (PIUNA Program) have supported this work. E. A. thanks the late M. {\'A}lvarez-Ponte for inspiration.
\end{acknowledgments}

\end{article}


\begin{thebibliography}{10}

\bibitem{Shinbrot2003} Shinbrot, T., Duong, N. -H., Kwan, L. and Alvarez, M. M. (2003) {\it PNAS} {\bf 101} 8542-8546.

\bibitem{Almeida2008} Almeida M. P., Parteli E. J., Andrade, Jr. J.
S. and Herrmann H. J. (2008) {\it PNAS} 6222-6226.

\bibitem{Thomas2005} Thomas, P. C. and Robinson, M. S. (2005) {\it
Nature} {\bf 436} 366-369.

\bibitem{Miyamoto2007} Miyamoto, H. {\it et al.} (2007) {\it
Science} {\bf 316} 1011-1014.

\bibitem{NASA2009} Jet Propulsion Laboratory (NASA) (2009) \\
 http://marsrover.nasa.gov/spotlight/20091019a.html

\bibitem{Elvis2012} Elvis, M. (2012) {\it Nature} {\bf 485} 549.

\bibitem{Uehara2003} Uehara J. S., Ambrosio M. A., Ohja R. P. and
Durian D. J. (2003) {\it Phys. Rev. Lett.} {\bf 90} 194301-194304.

\bibitem{Walsh2003} Walsh A. M., Holloway K. E., Habdas P. and de
Bruyn, J. R. (2003) {\it Phys. Rev. Lett.} {\bf 91} 104301-104304.

\bibitem{Boudet2006} Boudet J. F., Amarouchene Y. and Kellay H.
{\it Phys. Rev. Lett.} {\bf 96} 158001-158004.

\bibitem{Vet2007}  de Vet S. J. and Bruyn J. R. (2007) {\it Phys.
Rev. E} {\bf 76} 041306-041311.

\bibitem{Katsuragi2007}  Katsugari H. and Durian D. (2007){\it Nat.
Phys.} {\bf 3} 420-???

\bibitem{Pacheco2011} Pacheco-V{\'a}zquez F., Caballero-Robledo G.
A., Solano-Altamirano J. M., Altshuler E., Batista-Leyva A. J. and
Ruiz-Su{\'a}rez, J. C. (2011) {\it Phys. Rev. Lett.} {\bf 106}
218001-218004.

\bibitem{Katsuragi2012} Katsuragi H. (2012) {\it Phys. Rev. E} {\bf
85} 021301-021305.

\bibitem{Kondic2012} (2012) Kondic L., Fang X., Losert W., O'Hern C.
S. and Behringer R. P. {\it Phys. Rev. E} {\bf 85} 011305-011317.

\bibitem{Ruiz-Suarez2013} Ruiz-Su{\'a}rez J. C. (2013) Penetration of projectiles into
granular targets. {\it Rep Prog Phys} 76:~066601.

\bibitem{Goldman2008} (2008) Goldman, D. I. and Umbanhovar P. {\it
Phys. Rev. E} {\bf 77}, 021308-021311.

\bibitem{Brzinski2010} Brzinski T. A. and Durian D. J. (2010) {\it
Soft Matter} {\bf 6} 3038-3043.

\bibitem{Brzinski2013} T. A. Brzinski III, Mayor P., and Durian D. J., (2013)
{\it in preparation}.

\bibitem{Constantino2011} Constantino D. J., Bartell J., Scheidler
K. and Schiffer P. (2011) {\it Phys. Rev. E} {\bf 83} 011305-011308.

\bibitem{Li2009} Li Chen, Umbanhowar, P. B., Komsuoglu H.,
Koditschek, D. E and Goldman D. I. (2009) {\it PNAS} {\bf 106}
3029-3034.

\bibitem{Bi2011} Bi D., Zhang J., Chakraborty B. and Behringer R. P.
{\it Nature}, {\bf 480}, 355-358.

\bibitem{Torres2012} Torres H., Gonz{\'a}lez, A., S{\'a}nchez-Colina G., Drake J. C. and Altshuler E. (2012) {\it Rev. Cub. Fis.} {\bf 29} 1E45-1E47.

\bibitem{Janssen1895} Janssen H. A. (1895) {\it Z. Ver. Dt. Ing.} 39:1045.

\bibitem{Zigbee} See MMA7660FC ZSTAR3
accelerometer details at www.freescale.com/zstar.

\bibitem{Poschel2005} P\"{o}schel T. and Schwager T. (2005) {\em Computational Granular Dynamics}. Springer-Verlag, Berlin.

\bibitem{raul2013} R.C. Hidalgo T. Kanzaki,  F. Alonso-Marroquin, S. Luding (2013) {\em Proceedings of the Powders \& Grains 2013} to appear

\bibitem{sudafrica2013}  Juan-Pierre Longmore, Patrick Marais and Michelle Kuttel (2013) {\it Powder Technology}, {\bf 235}, pp 983-1000

\bibitem{Owens2008} Owens J., Houston M. Luebke D., Green S. Stone J. and Phillips J. (2008), ``Gpu
  computing,'' {\em Proceedings of the IEEE}, {\bf 96}, 879-899.

\bibitem{Fincham1992} Fincham D. (1992) ``Leapfrog rotational algorithms,'' {\it Molecular Simulation}, {\bf 8} 165-178.

\bibitem{Verlet1968} Verlet L. (1968)  {\it Phys. Rev.}, {\bf 165}, 201-214.


\end{thebibliography}
\end{document}